\documentclass[twocolumn,showpacs,preprintnumbers,amsmath,amssymb]{revtex4}
\usepackage{graphicx}

\begin{document}

\title{Quasiparticle spectrum and dynamical stability of an atomic Bose-Einstein
condensate coupled to a degenerate Fermi gas}
\author{C. P. Search, H. Pu, W. Zhang, and P. Meystre}
\affiliation{Optical Sciences Center, The University of Arizona,
Tucson, AZ 85721}
\date{\today}

\begin{abstract}
The quasiparticle excitations and dynamical stability of an atomic
Bose-Einstein condensate coupled to a quantum degenerate Fermi gas
of atoms at zero temperature is studied. The Fermi gas is assumed
to be either in the normal state or to have undergone a phase
transition to a superfluid state by forming Cooper pairs. The
quasiparticle excitations of the Bose-Einstein condensate exhibit
a dynamical instability due to a resonant exchange of energy and
momentum with quasiparticle excitations of the Fermi gas. The
stability regime for the bosons depends on whether the Fermi gas
is in the normal state or in the superfluid state. We show that
the energy gap in the quasiparticle spectrum for the superfluid
state stabilizes the low energy energy excitations of the
condensate. In the stable regime, we calculate the boson
quasiparticle spectrum, which is modified by the fluctuations in
the density of the Fermi gas.
\end{abstract}
\pacs{03.75.Fi,05.30.Fk}
\maketitle

\section{Introduction}
Since the observation of Bose-Einstein condensation in trapped
atomic gases \cite{BEC}, there has been increasing interest in
creating quantum degenerate gases of fermions with trapped
ultracold akali atoms. At temperatures below the Fermi
temperature, $T_{F}$, the properties of the gas become strongly
influenced by the Pauli exclusion principle \cite{Pauli,holland2}.
Besides exploring the role of quantum statistics in their
behavior, much of the interest in these gases has focused on the
possibility of achieving the Bardeen-Cooper-Schrieffer (BCS) phase
transition to the superfluid state by forming Cooper pairs
\cite{Li6,burnett, bohn,holland,mackie}.

Currently, experimental efforts in cooling of fermionic atoms of
$^{6}$Li \cite{truscott,thomas} and $^{40}$K \cite{jin,Pauli} to
the quantum degenerate regime have made significant progress,
reaching temperatures as low as $0.2T_{F}$ where $T_{F}$ is the
Fermi temperature. However, the efficiency of the evaporative
cooling process used to cool a two component Fermi gas is severely
hindered for temperatures below $T_{F}$ due to Pauli blocking
\cite{holland2}. Meanwhile, the lack of $s$-wave scattering
between spin-polarized fermions makes evaporative cooling
completely ineffective for a single component Fermi gas.
Furthermore, it has been recently predicted that loss processes
which remove particles from the trap and leave holes behind in the
single particle distribution also impose a lower limit on the
temperature ($\sim T_F/4$) that can be reached in a pure Fermi gas
\cite{eddy}. As a result, the recent experiments that achieved
quantum degeneracy in $^{6}$Li have used $^{7}$Li, a boson, to
sympathetically cool the $^{6}$Li atoms \cite{truscott}. This
procedure is also being applied to cool $^{40}$K using $^{87}$Rb
\cite{jin2}. Therefore, it appears likely that future experiments
on degenerate Fermi gases will be associated with a Bose gas with
a non-negligible boson-fermion two-body interaction. In a more
speculative vein, the nonlinear mixing of bosonic and fermionic
matter waves may open up the way to novel methods to manipulate
these waves, and in particular their statistical properties. The
theoretical study of the properties of coupled Bose-Einstein
condensates and degenerate Fermi gases is therefore of
considerable practical interest to future experiments.

It is the purpose of this paper to establish a general analysis of
the quasiparticle spectrum and dynamical stability for a BEC
coupled to a degenerate Fermi gas. There have been a number of
recent studies of both the ground state properties
\cite{molmer,viverit,roth,yi} and the collective modes for the
density fluctuations of the coupled gases
\cite{yip,bijlsma,capuzzi,minguzzi1}. These studies have all
treated the ground state of the fermions as being that of a normal
degenerate Fermi gas. The likelihood of observing the superfluid
state in fermionic akali vapors has sparked several theoretical
studies of the quasiparticle and collective modes of trapped
superfluid Fermi gases \cite{baranov,minguzzi,mottelson,bruun}.
However, the quasiparticle excitations of a {\it coupled}
superfluid Fermi gas and Bose-Einstein condensate (BEC) has not
yet been investigated. In this paper, we examine both the case
when the ground state of the fermions, in the absence of a
boson-fermion interaction, is the normal state and the BCS
superfluid state. In both cases, the quasiparticle spectrum
exhibits a dynamical instability due to the exchange of energy and
momentum with the fermions. Physically, the existence of the
instability implies the existence of a lower energy ground state
of the coupled system, which involves correlated density
fluctuations between the BEC and Fermi gas. In the stable regime,
the quasiparticle dispersion relation for the bosons is
significantly modified due to quantum fluctuations in the density
of the Fermi gas. More importantly, the stability regime and the
quasiparticle dispersion for the bosons is qualitatively different
when the fermions are in the BCS state as compared to the normal
state.

In terms of current experiments with trapped atomic gases, we are
only interested in the dilute limit for the Bose-Fermi mixture. In
that limit, we can linearize the equations of motion for the
density fluctuations of the two gases, since at zero temperature
the fluctuations relative to the non-interacting ground state are
expected to be small. For the bosons, this method is equivalent,
in the absence of a boson-fermion coupling, to the Bogoliubov
procedure \cite{bogo1}. The presence of a boson-fermion coupling
results in a modified boson-boson interaction due to the induced
density fluctuations in the Fermi gas. In Sec. II, we present our
model and derive the quasiparticle spectrum for the bosons when
the fermions are in the normal state. In Sec. III, we show how the
calculation of Sec. II is modified when the fermions are in the
BCS state. In the Appendix, we show how quantum correlations
between the densities of the two gases can lower the ground state
energy of the mixture.

\section{Mixture of BEC and Normal Fermi Gas}
This paper focuses on the effect that the boson-fermion
interaction has on the quasiparticles states of the Bose gas.
Hence, we neglect the effect of a direct interaction between the
fermions in this section. For a spin-polarized Fermi gas, this is
an excellent approximation since $s$-wave scattering between two
fermions is forbidden and $p$-wave scattering is negligible at
zero temperature. However, for the sake of generality, we consider
the case of fermions with two hyperfine spin states. The results
of this section can be directly applied to a single component
Fermi gas \cite{pu} since we assume that the boson-fermion
interaction and the single-particle energies of the fermions are
independent of the spin. In the next section, we will generalize
these results to the case of $s$-wave Cooper pairing. For this
purpose, it will be necessary to explicitly include an attractive
interaction in order to create a non-zero pairing field needed for
the BCS state.

Our starting point is the grand canonical Hamiltonian for a weakly
interacting gas of bosons coupled to an ideal gas of fermions with
the two spin states, labelled by $\sigma = \uparrow,\downarrow$,
\begin{equation}
\hat{H}=\hat{H}_{B}+\hat{H}_{F}+\hat{H}_{BF}, \label{H}
\end{equation}
where $\hat{H}_{B}$ and $\hat{H}_{F}$ are the free Hamiltonians
for the bosons and fermions, respectively,
\begin{eqnarray*}
\hat{H}_{B}&=& \int d^{3}r\, \hat{\psi}^{\dagger}_{B}({\bf r})
\left[
-\frac{\hbar^{2}\nabla^{2}}{2m_{B}} + V_{B}({\bf r}) -\mu_{B} \right.  \\
&&\left.+ \frac{g_{B}}{2}\hat{\psi}^{\dagger}_{B}({\bf r})
\hat{\psi}_{B}({\bf r}) \right] \hat{\psi}_{B}({\bf r}), \\
\hat{H}_{F}&=&\sum_{\sigma} \int d^{3}r\,
\hat{\psi}^{\dagger}_{\sigma}({\bf r}) \left[
-\frac{\hbar^{2}\nabla^{2}}{2m_{F}} + V_{F}({\bf r}) -\mu_{\sigma}
\right] \hat{\psi}_{\sigma}({\bf r}),
\end{eqnarray*}
while $\hat{H}_{BF}$ represents the boson-fermion interaction,
\begin{equation}
\hat{H}_{BF}= g_{BF}\sum_{\sigma} \int d^{3}r\,
\hat{\psi}^{\dagger}_{B}({\bf r})
\hat{\psi}^{\dagger}_{\sigma}({\bf r})\hat{\psi}_{\sigma}({\bf r})
\hat{\psi}_{B}({\bf r}). \nonumber
\end{equation}
Here, $\hat{\psi}_{B} ({\bf r})$ $[\hat{\psi}^{\dagger}_{B}({\bf
r})]$ and $\hat{\psi}_{\sigma}({\bf r})$ $
[\hat{\psi}^{\dagger}_{\sigma}({\bf r})]$ are the annihilation
(creation) operators for the bosons and for fermions with
hyperfine spin $\sigma$, respectively. They obey the standard
commutation (anti-commutation) relations. $\mu_{B}$ and
$\mu_{\sigma}$ are the chemical potentials for the bosons and
fermions. The coupling constants, $g_{B}$ and $g_{BF}$, are
defined as
\begin{eqnarray*}
g_B = 4\pi \hbar^2a_{B}/m_{B},\;\; g_{BF} = 2\pi \hbar^2
a_{BF}/m_{r},
\end{eqnarray*}
where $a_{B}>0$ and $a_{BF}$ are the boson-boson and boson-fermion
$s$-wave scattering lengths, respectively, while $m_r=m_B
m_F/(m_B+m_F)$ is the reduced mass. For simplicity, we assume that
the number of fermions in each spin state is the same so that
$\mu_{\uparrow,\downarrow}=\mu_F$.

To determine the excitation spectrum of the Bose-condensate, we
apply the standard Bogoliubov procedure by decomposing the field
operator as
\begin{equation}
\hat{\psi}_B ({\bf r},t) = \phi_B({\bf r}) + \hat{\xi}_B({\bf
r},t). \label{Bosewave}
\end{equation}
Here, $\hat{\xi}_B({\bf r},t)$ describes the small amplitude
fluctuations above the condensate mode, $\phi_B({\bf r})$, which
obeys the Gross-Pitaevskii equation,
\begin{eqnarray}
\left[ -\frac{\hbar^{2}\nabla^{2}}{2m_{B}} + V_{B}({\bf r})+
g_{B}|\phi_B({\bf r})|^2 + 2g_{BF}n_F({\bf r})-\mu_{B} \right]
\label{gross}
\\ \nonumber
\times \phi_B({\bf r})=0.
\end{eqnarray}
Similarly, we assume that the density fluctuations in the Fermi
gas are small so that
\begin{equation}
\hat{\psi}_{\sigma}^\dag ({\bf r},t) \hat{\psi}_{\sigma} ({\bf
r},t)= n_F({\bf r})+\delta \hat{\rho}_{\sigma}({\bf r},t)
\label{Fermdens}
\end{equation}
where $\langle\delta \hat{\rho}_{\sigma}({\bf r},t)\rangle=0$ in
the absence of any external perturbations. For a trapped gas, the
equilibrium density of each spin component, $n_F({\bf r})$, may be
approximated by the Thomas-Fermi expression for the density
\cite{molmer}. There is an obvious asymmetry in our treatment of
the bosons and fermions since for the BEC there is a nonvanishing
expectation of $\hat{\psi}_B ({\bf r},t)$ while for the fermions,
$\langle\hat{\psi}_{\sigma} ({\bf r},t)\rangle=0$ and only the
fluctuations in the fermion density may be regarded as being small
relative to some finite mean-field.

By substituting Eqs. (\ref{Bosewave}) and (\ref{Fermdens}) into
Eq.~(\ref{H}), and neglecting terms involving the product of three
or more fluctuation operators, one obtains a Hamiltonian that is
quadratic in $\hat{\xi}_B({\bf r},t)$ and
$\hat{\psi}_{\sigma}({\bf r})$. From this quadratic Hamiltonian,
one obtains Heisenberg equations of motion that are linear in the
fluctuations,
\begin{eqnarray}
i\hbar \frac{\partial \hat{\xi}_B}{\partial t} &=& \hat{h}_B
\hat{\xi}_B + g_{B}\phi_B^2 \hat{\xi}_B^\dagger +g_{BF}\phi_B
\sum_{\sigma} \delta \hat{\rho}_{\sigma} ,\label{b} \\
i\hbar \frac{\partial \hat{\psi}_{\sigma}}{\partial t} &=&
\hat{h}_F \hat{\psi}_{\sigma} + g_{BF}(\phi_B \hat{\xi}_B^\dagger
+ \phi_B^* \hat{\xi}_B) \hat{\psi}_{\sigma} ,\label{f}
\end{eqnarray}
where
\begin{eqnarray*}
\hat{h}_B&=&-\frac{\hbar^{2}\nabla^{2}}{2m_{B}} + V_{B}({\bf r})+
2g_{B}|\phi_B({\bf r})|^2 + 2g_{BF}n_F({\bf r})-\mu_{B}, \\
\hat{h}_F&=&-\frac{\hbar^{2}\nabla^{2}}{2m_{F}}+V_{F}({\bf
r})+g_{BF}|\phi_B({\bf r})|^2-\mu_F.
\end{eqnarray*}
Equations (\ref{b}) and (\ref{f}) can be thought of as describing
a four-wave mixing process where a bosonic wave, $\phi_B({\bf
r})$, scatters off the fermionic density grating, $\delta
\hat{\rho}_{\sigma}$, to create a new bosonic wave,
$\hat{{\xi}_B}({\bf r},t)$. Equation (\ref{f}) represents the
back-action on the fermion grating as a result of the scattering
of the bosonic wave. In contrast to Ref.\cite{moore}, where the
fermionic grating was created optically, or matter-wave
superradiance \cite{superrad}, where the grating results from the
mixing of optical and matter waves, four-wave mixing results now
from the coupling between the bosonic and fermionic matter-wave
fields.

The procedure we adopt here is to formally integrate Eq.~(\ref{f})
to obtain a linearized expression for $\delta
\hat{\rho}_{\sigma}$, which can then be substituted back into Eq.
(\ref{b}) to obtain an integro-differential equation for the boson
fluctuation operators. To begin, we expand the fermion field
operators in terms of the eigenstates of $\hat{h}_F$,
\begin{equation}
\hat{\psi}_{\sigma} ({\bf r},t)=\sum_n\,\hat{a}_{n,\sigma}(t)
\varphi_n({\bf r}), \label{an}
\end{equation}
where $\hat{h}_F \varphi_n({\bf r})=E_n \varphi_n({\bf r})$. The
formal solution of Eq.~(\ref{f}) is then given by
\begin{eqnarray}
\hat{\psi}_{\sigma} ({\bf r},t)&=& \hat{\psi}_{\sigma}^{(0)}({\bf
r},t)-i\frac{g_{BF}}{\hbar}
\int_0^t dt'\int d^3r'\,G({\bf r},{\bf r'},t-t') \nonumber \\
&&\;\; \times \,\Xi ({\bf r'},t') \hat{\psi}_{\sigma}({\bf r'},t')
, \label{psi}
\end{eqnarray}
where
\begin{equation}
\hat{\psi}_{\sigma}^{(0)}({\bf r},t) = \sum_n
\hat{a}_{n,\sigma}(0) e^{-iE_n t/\hbar} \varphi_n({\bf r})
\end{equation}
represents the free evolution of the field in the absence of a
density fluctuation of the BEC, and
\begin{equation}
\Xi ({\bf r},t) = \phi_B({\bf r}) \hat{\xi}_B^\dag({\bf r},t) +
\phi_B^*({\bf r}) \hat{\xi}_B ({\bf r},t).
\end{equation}
Physically, $\Xi ({\bf r},t)$ produces a density grating off which
the fermions can scatter. Consequently, the second term on the
right hand side of Eq. (\ref{psi}) may be interpreted as the
scattering of the fermions off the potential $g_{BF} \Xi ({\bf
r'},t')$ and the subsequent propagation of the fermions to $({\bf
r},t)$ by the single particle Green's function,
\begin{equation}
G({\bf r},{\bf r'},t-t') \equiv \sum_n e^{-iE_n (t-t')/\hbar}
\varphi_n({\bf r})\varphi_n^*({\bf r'}).
\end{equation}
In order to obtain a linear equation for the boson fluctuations,
we make the first Born approximation in Eq.~(\ref{psi}), so that
\begin{eqnarray}
\hat{\psi}_{\sigma} ({\bf r},t) &\approx&
\hat{\psi}_{\sigma}^{(0)}({\bf r},t)-i\frac{g_{BF}}{\hbar}
\int_0^t dt'\int d^3r'\,G({\bf r},{\bf r'},t-t') \nonumber \\
&&\;\;\times \,\Xi ({\bf r'},t') \hat{\psi}_{\sigma}^{(0)}({\bf
r'},t') . \label{psi1}
\end{eqnarray}
An expression for the fermion density fluctuation that is linear
in the boson fluctuations is obtained from Eq.~(\ref{psi1}) and
\begin{eqnarray}
\delta \hat{\rho}_{\sigma} =\langle F| \hat{\psi}_{\sigma}^\dagger
({\bf r},t) \hat{\psi}_{\sigma} ({\bf r},t)|F\rangle - n_F({\bf
r}) ,\label{rho1}
\end{eqnarray}
where $|F\rangle$ represents the zero temperature ground state of
the Fermi gas. By making use of the fact that at $T=0$, $\langle
F| \hat{a}^{\dagger}_{n,\sigma}(0)\hat{a}^{}_{n',\sigma '}(0)
|F\rangle = \delta_{n,n'}\delta_{\sigma,\sigma'}$ for $E_n\leq
E_F$ and zero otherwise ($E_F$ is the Fermi energy) as well as
$\sum_n |\varphi_n({\bf r})|^2=n_F({\bf r})$, we obtain the
desired expression for the fermion density fluctuation due to the
coupling to the bosons,
\begin{equation}
\delta \hat{\rho}_{\sigma} \approx i\frac{g_{BF}}{\hbar} \int_0^t
dt' \int d^3r'\, {\cal J} ({\bf r},{\bf r'},t-t') \Xi ({\bf
r'},t') , \label{deltaF}
\end{equation}
where
\begin{eqnarray*}
{\cal J}({\bf r},{\bf r'},t-t') &\equiv& G^*_>({\bf r},{\bf
r'},t-t') G_< ({\bf
r},{\bf r'},t-t') \\
&&\;-G_>({\bf r},{\bf r'},t-t') G_<^*({\bf r},{\bf r'},t-t') ,\\
G_>({\bf r},{\bf r'},t-t') &\equiv& \sum_{\{ n|E_n > E_F \}}
e^{-iE_n(t-t')/\hbar}  \varphi^*_n({\bf r'}) \varphi_n({\bf r}),
\end{eqnarray*}
and $G_<$ is the same as $G_>$, but for $E_n \le E_F$ in the
summation. By inserting Eq.~(\ref{deltaF}) back into Eq.~(\ref{b})
we finally have
\begin{widetext}
\begin{eqnarray}
i\hbar \frac{\partial \hat{\xi}_B ({\bf r},t)}{\partial t}
=\hat{h}_B \hat{\xi}_B ({\bf r},t) + g_{B}\phi_B^2
\hat{\xi}_B^\dagger ({\bf r},t)  + i\frac{2g_{BF}^2 }{\hbar}
\int_0^t dt' \int d^3r'\, {\cal J} ({\bf r},{\bf r'},t-t')  \Xi
({\bf r'},t') \phi_B({\bf r}). \label{bb}
\end{eqnarray}
\end{widetext}
Equation~(\ref{bb}) is valid to all orders in $g_{BF}$ provided
the density fluctuations of the bosons, $\Xi ({\bf r},t)$, and the
fermions, $\delta \hat{\rho}_{\sigma}({\bf r},t)$, remain small
relative to the equilibrium densities of the two gases. The
dependence of $\hat{\xi}_B$ to all orders in $g_{BF}$ is easily
seen by iterating the expression for $\Xi ({\bf r},t)$ inside the
integrand to obtain a power series expansion in even powers of
$g_{BF}$.

The physical interpretation of the integral term in Eq.~(\ref{bb})
is straightforward. Since the boson-fermion interaction is
proportional to the local densities of the two gases, a density
fluctuation in the BEC at ${\bf r'}$ and $t'$, $\Xi ({\bf
r'},t')$, will excite a density fluctuation in the Fermi gas at
the same point. This density fluctuation consists of particles
excited above the Fermi surface, which are represented by $G_>$,
and holes inside the Fermi sea, represented by $G_<$. These
particle-hole pairs then propagate from ${\bf r'}$ and $t'$ to
${\bf r}$ and $t$ where they excite another fluctuation in the
density of the BEC, thereby modifying the value of $\hat{\xi}_B
({\bf r},t)$ and hence, $\Xi ({\bf r},t)$. The integral in Eq.
(\ref{bb}) may then be thought of as a feedback loop where the
density fluctuations of the Fermi gas act as the feedback
mechanism. If the BEC is stable, the feedback will be negative
while for an unstable system, the coupled boson-fermion density
fluctuations will result in a positive feedback, which causes the
fluctuations to grow exponentially in time.

Equation (\ref{bb}), which is the main result of this section, is
completely general. However, for concreteness, we consider a
homogenous mixture of $N_B$ bosons and $2N_F$ fermions confined in
a box of volume $V$.  The corresponding ground state densities are
$n_B=|\phi({\bf r})|^2=N_B/V$ and $n_F=n_F ({\bf r})=N_F/V$. The
chemical potentials are $\mu_B=g_B n_B+2g_{BF} n_F$ [see
Eq.~(\ref{gross})] and $E_F=\hbar^2 k_F^2/2m_F=\mu_F-g_{BF}n_B$
where the Fermi wave number is given by $k_F=[(6\pi^2)n_F]^{1/3}$.
For periodic boundary conditions, the eigenstates of $\hat{h}_F$
are plane waves $\varphi_{{\bf k}} ({\bf r})=e^{i{\bf k\cdot
r}}/\sqrt {V}$ with eigenenergies $E_{{\bf k}}=\hbar^2
k^2/2m_F-E_F$. Similarly, we use a plane wave basis for the boson
field operator,
\begin{equation}
\hat{\psi}_B ({\bf r},t)=\frac{1}{\sqrt{V}} \sum_{{\bf k}}
\eta_{{\bf k}}(t) e^{i {\bf k}\cdot {\bf r}}\,.
\end{equation}
Hence, the boson fluctuation operator consists of all modes with
${\bf k}\neq 0$,
\begin{equation}
\hat{\xi}_B ({\bf r},t)=\frac{1}{\sqrt{V}} \sum_{{\bf k}\neq 0}
\eta_{{\bf k}}(t) e^{i {\bf k}\cdot {\bf r}}\,.
\end{equation}
The quantum state of the mixture can therefore be expressed as
\begin{equation}
|\Psi_N\rangle=\frac{1}{\sqrt{N_B!}}\left(\eta^{\dagger}_0\right)^{N_B}
\prod_{k\leq k_F,\sigma}a^{\dagger}_{{\bf k}\sigma}|0\rangle,
\label{Psi_N}
\end{equation}
where $|0\rangle$ represents the vacuum state for both the bosons
and fermions. Note that $|\Psi_N\rangle$ is identical to the state
assumed by Viverit {\it et al.} \cite{viverit}, if one includes
two spin components.

The function ${\cal J}({\bf r},{\bf r'},t-t')$, which describes
the propagation of particle-hole pairs in the Fermi gas, can now
be expressed explicitly as
\begin{eqnarray*}
{\cal J}({\bf r},{\bf r'},t-t')=&\frac{1}{V^2} &\sum_{k_n>k_F}
\sum_{k_m \le k_F} \left[
e^{i\frac{\hbar}{2m_F} (k_n^2-k_m^2)(t-t')}\right. \\
&& \times \left. e^{i({\bf k_m}-{\bf k_n}) \cdot {\bf r}}
 e^{-i({\bf k_m}-{\bf k_n}) \cdot {\bf r'}} -c.c. \right] .
\end{eqnarray*}
Multiplying Eq.~(\ref{bb}) by $e^{-i{\bf k \cdot r}}/\sqrt{V}$ and
integrating over ${\bf r}$ one obtains
\begin{equation}
i\hbar \frac{\partial \eta_{{\bf k}}}{\partial t} = {\cal L}_B
(k)\eta_{{\bf k}}+ g_Bn_B \eta_{-{\bf k}}^\dag +\frac{i}{\hbar}
g_{BF}^2 n_B I({\bf k}) ,\label{eta2}
\end{equation}
where ${\cal L}_B (k)=\hbar^2 k^2/2m_B+g_Bn_B$, and
\begin{widetext}
\begin{eqnarray*}
I({\bf k}) &=& \frac{2}{V}\int_0^t dt' \int d^3r' \int d^3r
\,e^{-i {\bf k}\cdot {\bf r}} {\cal J}({\bf r},{\bf r'},t-t')
 \,\sum_{{\bf k'}} e^{i {\bf k'}\cdot {\bf r'}} \left[
\eta^\dagger_{-{\bf k'}} (t') + \eta_{{\bf k'}}(t')
\right] \\
&=&\frac{2}{V}\int_0^t dt'\, \sum_{ k_m \le k_F}
\left[e^{i\frac{\hbar}{2m_F}  (|{\bf k_m} + {\bf k}|^2-k_m^2)
(t-t')}-c.c. \right]   \, \Theta (|{\bf k_m} + {\bf k}|-k_F)
\left[ \eta^\dagger_{-{\bf k}} (t') + \eta_{{\bf k}}(t') \right]
,
\end{eqnarray*}
\end{widetext}
where $\Theta(x)$ is the unit step function. By taking the adjoint
of Eq.~(\ref{eta2}) and using $I^{\dagger}(-{\bf k})=-I({\bf k})$
one gets,
\begin{equation}
i\hbar \frac{\partial \eta^\dag_{-{\bf k}}}{\partial t} = -{\cal
L}_B (k)\eta^\dag_{-{\bf k}}- g_Bn_B \eta_{{\bf k}}
-\frac{i}{\hbar} g_{BF}^2 n_B I({\bf k}) .\label{eta2+}
\end{equation}

The coupled integro-differential Eqs.~(\ref{eta2}) and
(\ref{eta2+}) may be solved using Laplace transforms. Denoting the
single-sided Laplace transforms as,
\begin{eqnarray*}
\alpha_{{\bf k}}(s) &=& L[\eta_{{\bf k}}(t)] =\int^{\infty}_0
dt\,e^{-st} \eta_{{\bf k}}(t) ,\\
\beta_{{\bf k}}(s) &=& L[\eta^\dagger_{-{\bf k}}(t)]=
\int^{\infty}_0 dt\,e^{-st} \eta^{\dagger}_{-{\bf k}}(t),
\end{eqnarray*}
one obtains the inhomogeneous linear equations
\begin{widetext}
\begin{subequations}
\begin{eqnarray}
i\hbar [s \alpha_{{\bf k}}(s)-\eta_{{\bf k}}(0)] &=& {\cal L}_B(k)
\alpha_{{\bf k}}(s) +g_Bn_B \beta_{{\bf k}}(s)
+\frac{g_{BF}^2}{\hbar (2\pi)^3}n_B \ell_k(s)[\alpha_{{\bf k}}(s)+
\beta_{{\bf k}}(s)] ,  \\
i\hbar [s\beta_{{\bf k}}(s)-\eta^\dagger_{-{\bf k}}(0)] &=& -{\cal
L}_B(k)\beta_{{\bf k}}(s) -g_Bn_B \alpha_{{\bf k}}(s)   -
\frac{g_{BF}^2}{\hbar (2\pi)^3}n_B \ell_k(s) [\alpha_{{\bf k}}(s)+
\beta_{{\bf k}}(s)],
\end{eqnarray}
\label{s}
\end{subequations}
where
\begin{eqnarray}
\ell_k(s) &=& \frac{2i(2\pi)^3}{V} L\left[\sum_{k_m \le
k_F}^{|{\bf k_m}+{\bf k}| > k_F} e^{i(|E_{{\bf k+k_m}}|+|E_{{\bf
k_m}}|)t/\hbar}
-c.c. \right]  \label{dens-dens} \\
&=& 2 \int d{\bf k_m} \left[ \frac{1}{-is-(|E_{{\bf
k+k_m}}|+|E_{{\bf k_m}}|)/\hbar}  -\frac{1}{-is+(|E_{{\bf
k+k_m}}|+|E_{{\bf k_m}}|)/\hbar} \right]\Theta(|{\bf k_m}+{\bf
k}|-k_F) \Theta(k_F-k_m).\nonumber
\end{eqnarray}
\end{widetext}
and in the second line we have taken the infinite volume limit to
convert the summation to an integral. The solutions of
Eqs.~(\ref{s}) are,
\begin{subequations}
\begin{eqnarray}
\alpha_{{\bf k}}(s)&=&i\hbar \frac{\left[\zeta_k(s) +\left(i\hbar
s+T_k \right)\eta_{{\bf k}}(0) \right]}{(i\hbar
s)^2-z_k(s)}\,, \\
\beta_{{\bf k}}(s)&=&i\hbar \frac{\left[-\zeta_k(s)+\left(i\hbar
s-T_k \right)\eta^{\dagger}_{-{\bf k}}(0) \right]}{(i\hbar
s)^2-z_k(s)}\,,
\end{eqnarray}
\label{as}
\end{subequations}
where $T_k=\hbar^2k^2/(2m_B)$ and
\begin{eqnarray*}
\zeta_k(s)&=&n_B\left[g_B+\frac{g_{BF}^2 \ell_k(s)}{\hbar
(2\pi)^3} \right] \left[\eta_{{\bf k}}(0)+\eta^{\dagger}_{-{\bf
k}}(0)\right], \\
z_k(s)&=&T_k\left[T_k+2n_B \left( g_B + \frac{g_{BF}^2}{\hbar
(2\pi)^3} \ell_k(s) \right) \right].
\end{eqnarray*}
The poles of Eqs.~(\ref{as}) in the $s$-plane correspond to the
quasiparticle excitation frequencies of the condensate. For
$g_{BF}=0$, one obviously recovers the Bogoliubov spectrum of a
pure weakly interacting BEC. For $s=i\omega+0^+$, $\ell_k(s)$ is
proportional to the density-density response function of an ideal
Fermi gas, which measures the linear response of the density of
the gas to a scalar perturbing potential \cite{pines}. The
expression $g_B + g_{BF}^2\ell_k(s)/\hbar (2\pi)^3$ corresponds to
a renormalized boson-boson interaction due to the polarization of
the density of the Fermi gas.

We already mentioned that in case of positive feedback between the
density fluctuations in the two gases, the BEC becomes unstable.
Mathematically, the stability of the BEC is determined by the
location of the poles in the $s$-plane,
\begin{equation}
(i\hbar s)^2-z_k(s)=0. \label{poles}
\end{equation}
In order for the BEC to be stable, Re$[s]<0$ for all solutions of
(\ref{poles}). A positive real part of any of the poles indicates
the existence of an instability in the BEC.

To evaluate the boson excitation frequencies in the stable regime
as well as the location of the instabilities it is sufficient to
make the substitution $s=i\omega+0^+$, in which case the condition
that the BEC be stable corresponds to poles with Im$[\omega]>0$.
The real and imaginary parts of
$\ell_k(\omega)\equiv\ell_k(s=i\omega+0^+)$ can be obtained using
$1/(x\pm i0^{+})=P (1/x) \mp i\pi\delta(x)$ \cite{fetter},
\begin{widetext}
\begin{eqnarray}
{\rm Re} [\ell_{k}(\omega)]& =& \frac{2\pi}{\alpha}\left\{ \left[
k_F^2- \left(\frac{\omega}{\alpha}-\frac{k}{2}\right)^2 \right]
\ln \left| \frac{\omega+\alpha(k_F-k/2)}{\omega-\alpha(k_F+k/2)}
\right| +\left[k_F^2-
\left(\frac{\omega}{\alpha}+\frac{k}{2}\right)^2 \right] \ln
\left| \frac{\omega-\alpha(k_F-k/2)}{\omega+\alpha(k_F+k/2)}
\right|-2k_F k \right\}
,\nonumber \\
{\rm Im} [\ell_{k}(\omega)] &= & 2\pi \int d{\bf k_m}
\,\Theta(|{\bf k_m}+{\bf k}|-k_F) \Theta(k_F-k_m) \times
\,\delta\left( \omega-|E_{{\bf k_m}+{\bf k}}|/\hbar- |E_{{\bf
k_m}}|/\hbar \right),
 \label{im}
\end{eqnarray}
\end{widetext}
where $\alpha=\hbar k/m_F$.

The mechanical stability of the condensate, which requires its
compressibility to be positive, can be derived from the zero
frequency ($\omega = 0$) static limit for the speed of sound in
the condensate. From Eq.~(\ref{poles}), in the long wavelength
limit (i.e., $k \rightarrow 0$), we have $\omega^2=c_B^2 k^2$
where $c_B$ is the speed of sound. A positive compressibility
corresponds to $c_B^2\geq 0$. From Eq.~(\ref{poles}), $c_B^2$ is
given by,
\begin{equation}
c_B^2=\frac{n_B}{m_B}\left[ g_B + \frac{g_{BF}^2}{\hbar (2\pi)^3}
\lim_{k\rightarrow 0} \ell_{k}(0)\right].
\end{equation}
Using the expansion of $\ell_{k}(0)$ for $k\ll k_F$,
\begin{equation}
\ell_{k}(0)\approx\frac{4\pi
m_Fk_F}{\hbar}\left[-2+\frac{1}{6}\left(\frac{k}{k_F}\right)^2
\right] ,\label{limit}
\end{equation}
one obtains the mechanical stability condition,
\begin{equation}
n_F^{1/3}\leq \frac{A}{3} \frac{g_B}{g_{BF}^2} \,,\label{mechstab}
\end{equation}
where $A=(\hbar^2/2m_F)(6\pi^2)^{2/3}$, or in terms of scattering
lengths, \[a_{BF}^2 \leq \frac{\pi a_Bm_Bm_F}{k_F(m_B+m_F)^2}.\]
Eq.~(\ref{mechstab}) agrees with the result obtained in Refs.
\cite{viverit,yip,bijlsma} when one accounts for the two spin
components. In the following it will be assumed that Eq.
(\ref{mechstab}) is satisfied.

A necessary condition for the BEC to be {\it dynamically} stable
with respect to density fluctuations of finite energy and momentum
is that ${\rm Im}[\ell_{k}(\omega)]=0$. Since ${\rm
Im}[\ell_{k}(\omega)]$ is proportional to the dynamic structure
factor of the Fermi gas, it measures the rate at which energy and
momentum can be resonantly transferred between the density
fluctuations in the Fermi and Bose gases \cite{pines}. In the
absence of any decay or dephasing mechanism for the boson or
fermion quasiparticles, the existence of a bosonic quasiparticle
with energy $\hbar {\rm Re}[\omega ({\bf k})]$ such that ${\rm
Im}[\ell_{k}(\omega)]\neq 0$ will give rise to coupled
oscillations between the density fluctuations with wave vector
${\bf k}$, similar to Rabi oscillations for a two-level atom
coupled to a quantized field \cite{meystre}. One may then consider
a new ground state, which is a superposition of a bosonic and
fermionic density fluctuation, in analogy to the dressed states of
quantum optics \cite{meystre}. In the appendix it is shown that
such a superposition can result in a state with an energy lower
than the state used in this section and in other studies of
Bose-Fermi mixtures
\cite{molmer,viverit,roth,yi,yip,bijlsma,capuzzi,minguzzi}, which
do not contain any quantum correlations between the densities of
the two gases. This indicates that a dynamical instability
signified by ${\rm Im}[\ell_{k}(\omega)]\neq 0$ leads to a lower
energy ground state of the Bose-Fermi mixture. It is worth noting
that the dynamical instability is distinct from the mechanical
instability of the mixture, discussed previously. A mechanical
instability due to the Bose-Fermi coupling leads to a demixing of
the two gases and occurs in the static ($\omega=0$) limit for
which ${\rm Im}[\ell_{k}(\omega)]$ is always zero.

For excitations of the BEC with frequency $\omega$ and wave number
$k < 2k_F$, the stability criterion determined by ${\rm Im}
[\ell_{k}(\omega)] =0$ requires the excitation frequency to
satisfy
\begin{equation} \omega > \frac{\hbar k^2}{2m_F} +
\frac{\hbar k k_F}{m_F}  . \label{cri} \end{equation} Physically,
$\hbar^2 k^2/2m_F+\hbar^2 k k_F/m_F$ is the maximum energy that a
particle-hole pair can have for a given $k$. Hence, the stability
criterion corresponds to there being no excitations of the Fermi
gas that can resonantly couple to the condensate quasiparticle.
\begin{figure}
\begin{center}
    \includegraphics*[width=0.95\columnwidth,
height=0.7\columnwidth]{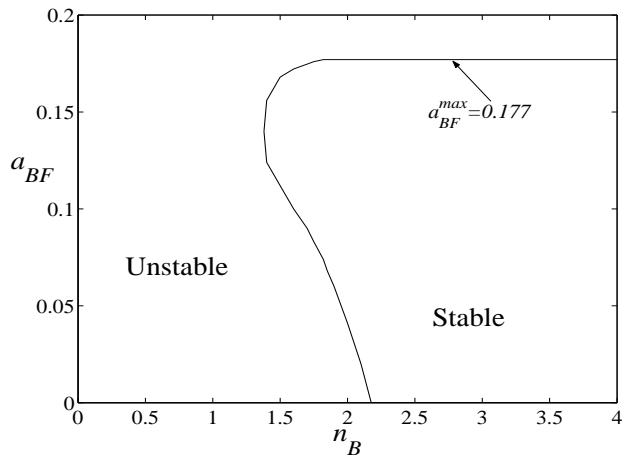} \vspace{3 mm} \caption{Stability
diagram of a Bose-Fermi mixture. We assume $m_F=m_B=m$. We have
adopted a system of units in which the units for frequency,
length, and wavenumber are $\hbar k_F^2/2m$, $1/k_F$, and $k_F$,
respectively. Note that in this units, the fermion density is
given by $n_F=1/(6\pi^2) \approx 0.017$.} \label{fig1}
\end{center}
\end{figure}
The stability regime and the phonon spectrum can be obtained by
first solving
\begin{eqnarray}
(\hbar\omega)^2 &=& z_k(i\omega+0^+) \nonumber \\
&=& T_k\left[T_k+2n_B \left( g_B + \frac{g_{BF}^2}{\hbar
(2\pi)^3}{\rm Re} [ \ell_k(\omega)] \right)\right],
\label{dispersion}
\end{eqnarray}
numerically while assuming Im$[\ell_{k}(\omega)]=0$, and then
checking if the system is stable using the criterion (\ref{cri}).
Fig.~\ref{fig1} shows the stability diagram of the mixture in the
$a_{BF}-n_B$ space. As can be seen, the dynamical stability of the
system is determined by both the scattering lengths and the atomic
densities. All other parameters being fixed, the stability
condition (\ref{cri}) imposes a minimum boson density $n_B^{min}$
beyond which no stable homogeneous mixture exists. For $k\ll k_F$,
we can use the linear part of the Bogoliubov spectrum for a pure
condensate to estimate $n_B^{min}$ as   \[g n_B^{min}\approx
\frac{\hbar^2 k_F^2}{m_F^2/m_B} =\frac{\hbar^2 (6\pi^2
n_F)^{2/3}}{m_F^2/m_B}.\] For realistic numbers, $n_B^{min}$ is
about two orders of magnitude larger than $n_F$ (see
Fig.~\ref{fig1}).
\begin{figure}
\begin{center}
    \includegraphics*[width=0.95\columnwidth,
height=0.7\columnwidth]{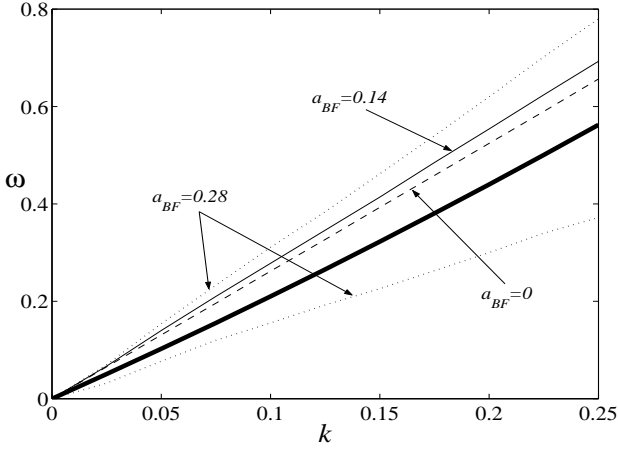} \vspace{3 mm} \caption{Phonon
spectrum of a boson-fermion mixture. The thick solid line
corresponds to $\omega=\hbar(k^2+2kk_F)/2m_F$. Frequencies fall
below this line represent unstable excitations. Same units as in
Fig.~\ref{fig1}.} \label{fig2}
\end{center}
\end{figure}

Figure~\ref{fig2} illustrates the phonon spectrum for the BEC in
the Bose-Fermi mixture when $n_B>n_B^{min}$.  The boson-fermion
interaction increases the sound velocity of the phonons. This
effect has a straightforward explanation in terms of the stability
condition imposed on the boson quasiparticle excitation frequency.
In the stable regime determined by (\ref{cri}), the
density-density response function $\ell_{k}(\omega)$ can be easily
shown to be positive, hence $\omega(k)$, as given by Eq.
(\ref{dispersion}), is increased.

For small values of $a_{BF}$, the spectrum is stable and
single-valued.  By expanding $\ell_{k}(\omega)$ around small
$k/k_F$ and finite $\omega$, we can calculate the sound velocity,
$c$, in the condensate for finite $\omega$ and $k$. Expanding
$\ell_{k}(\omega)$ to lowest order in $k/k_F$, we obtain
$\ell_{k}(\omega)=(8\pi\hbar /3)(k_F^3k^2/m_F\omega^2)$, and hence
the sound velocity is
\begin{equation}
c \approx c_0 \left[1+\frac{g_{BF}^2k_F^3}{3\pi^2g_Bm_Fc_0^2}
\right], \label{sound}
\end{equation}
where $c_0=\sqrt{g_Bn_B/m_B}$ is the sound velocity in a pure
condensate. For the parameters in Fig.~\ref{fig2}, we have a 6\%
increase in sound velocity if $a_{BF}$ changes from 0 to 0.14.
However, further increasing $a_{BF}$ beyond a critical value
$a_{BF}^{max}$ splits the phonon spectrum into two branches, and
one of them falls into the unstable regime. The critical value of
$a_{BF}$ for which the spectrum splits into two branches
corresponds to the equality in Eq. (\ref{mechstab}). For $m_F
\approx m_B$, we have $a_{BF}^{max}=\sqrt{\pi a_B/k_F}/2$, which
gives the maximum sound velocity achievable in a homogeneous
mixture
\[ c_{max} = c_0 \left( 1+\frac{c_F^2}{c_0^2} \right), \]
where $c_F=\hbar k_F/(\sqrt{3}m_F)$ is sound velocity of the ideal
Fermi gas \cite{fetter}. Note that while $n_B^{min}$ is
independent of $a_{BF}$, $a_{BF}^{max}$ is independent of $n_B$.

\section{Mixture of BEC and Superfluid Fermi Gas}
In order for a BCS transition to occur in a system of fermions,
there must be an attractive two-body interaction which allows the
fermions to form Cooper pairs. At the ultracold temperatures
achieved in current experiments, $p$-wave collisions between atoms
are highly suppressed and $s$-wave collisions between atoms in the
same internal state are forbidden by the Pauli exclusion
principle. As a result, the most likely possibility for the
formation of Cooper pairs is an attractive $s$-wave interaction
between atoms in different hyperfine states. Fortunately, $^{6}$Li
and $^{40}$K appear to be very promising candidates. $^{6}$Li
possesses an anomalously large and negative $s$-wave scattering
length, $a=-2160a_{B}$ where $a_{B}$ is the Bohr radius
\cite{Li6}. For $^{40}$K, a Feshbach resonance exists for two of
the hyperfine states which can be used to create a large negative
scattering length of $a\approx -1000a_{B}$ \cite{bohn}. To deal
with this situation, we now include in Eq. (\ref{H}) the term
\begin{equation}
\hat{H}_{FF}=-g_F\int d^3r\, \hat{\psi}^{\dagger}_{\uparrow}({\bf
r})\hat{\psi}^{\dagger}_{\downarrow}({\bf
r})\hat{\psi}_{\downarrow}({\bf r})\hat{\psi}_{\uparrow}({\bf r}),
\end{equation}
where $g_F=4\pi \hbar^2|a_F|/m_F$ and $a_F<0$ is the $s$-wave
scattering length between fermions in the spin singlet state.
$\hat{H}_{FF}$ can be treated using the self-consistent field
method by replacing pairs of fermion operators in $\hat{H}_{FF}$
with $c$-numbers \cite{degennes},
\begin{eqnarray}
\hat{H}_{FF}&=&\int d^3r\left[-g_Fn_F({\bf
r})\sum_{\sigma}\hat{\psi}^{\dagger}_{\sigma}({\bf
r})\hat{\psi}_{\sigma}({\bf r}) \right. \nonumber \\
&&\left. + \Delta({\bf r})\hat{\psi}^{\dagger}_{\uparrow}({\bf
r})\hat{\psi}^{\dagger}_{\downarrow}({\bf r})+\Delta^*({\bf
r})\hat{\psi}_{\downarrow}({\bf r})\hat{\psi}_{\uparrow}({\bf r})
\right], \label{pairing}
\end{eqnarray}
where
\begin{equation}
\Delta({\bf r})=-g_F\langle\hat{\psi}_{\downarrow}({\bf
r})\hat{\psi}_{\uparrow}({\bf r})\rangle,
\end{equation}
and the expectation value is taken with respect to the BCS ground
state defined below. $\Delta({\bf r})$ is the order parameter for
the BCS state. It represents the correlation between fermions that
have formed Cooper pairs and is zero for the normal state of the
Fermi gas. The term proportional to $-g_Fn_F({\bf r})$, is a
Hartree-Fock mean-field which is present even in the normal state
for an interacting Fermi gas. The inclusion of a Hartree-Fock term
for the normal state does not affect any of the results of the
last section since, in particular for a uniform system, it can be
absorbed into the definition of the Fermi energy.

We now proceed as before and calculate the quasiparticle spectrum
for the bosons. Eq.~(\ref{b}) is still valid, but Eq.~(\ref{f}) is
now replaced by the pair of equations
\begin{subequations}
\begin{eqnarray}
i\hbar \frac{\partial \hat{\psi}_{\uparrow}}{\partial t} &=
\hat{h}_F'\hat{\psi}_{\uparrow} + \Delta({\bf
r})\hat{\psi}^{\dagger}_{\downarrow}+g_{BF}\Xi ({\bf
r},t) \hat{\psi}_{\uparrow},  \\
i\hbar \frac{\partial \hat{\psi}^{\dagger}_{\downarrow}}{\partial
t} &= -\hat{h}_F'\hat{\psi}^{\dagger}_{\downarrow} + \Delta^*({\bf
r})\hat{\psi}_{\uparrow}-g_{BF}\Xi ({\bf r},t)
\hat{\psi}^{\dagger}_{\downarrow} ,
\end{eqnarray}
\label{fBCS1}
\end{subequations}
where $\hat{h}_F'=\hat{h}_F-g_Fn_F({\bf r})$. In the absence of
the Bose-Fermi coupling, Eqs.~(\ref{fBCS1}) may be solved by a
canonical transformation \cite{bogo2},
\begin{subequations}
\begin{eqnarray}
\hat{\psi}_{\uparrow}({\bf r},t)&=&\sum_n \left[u_n({\bf
r})\alpha_{n \uparrow}(t)-v_n^*({\bf r})\alpha_{n
\downarrow}^{\dagger}(t)\right],
 \\
\hat{\psi}^{\dagger}_{\downarrow}({\bf r},t)&=&\sum_n
\left[u_n^*({\bf r})\alpha_{n \downarrow}^{\dagger}(t)+v_n({\bf
r})\alpha_{n \uparrow}(t)\right],
\end{eqnarray}
\label{can1} \end{subequations}
where $\alpha_{n \sigma}$ $
(\alpha_{n \sigma}^{\dagger})$ are annihilation (creation)
operators for quasiparticles, which obey fermionic anticommutation
relations $\{\alpha_{n \sigma}^{\dagger},\alpha_{m \sigma '}\}=
\delta_{n,m}\delta_{\sigma,\sigma '}$ and $\{\alpha_{n
\sigma},\alpha_{m \sigma '}\}=0$. As a result, the amplitudes,
$u_n$ and $v_n$, are subject to the orthonormality condition
\begin{equation}
\int d^3r \,\left[u_n^*({\bf r})u_m({\bf r})+v_n^*({\bf
r})v_m({\bf r}) \right]=\delta_{n,m}. \nonumber
\end{equation}
The formal solutions of Eqs.~(\ref{fBCS1}) in terms of the
quasiparticle operators are
\begin{widetext}
\begin{subequations}
\begin{eqnarray}
\alpha_{n \uparrow}(t)&=&\alpha_{n
\uparrow}(0)e^{-i\varepsilon_nt/\hbar} \nonumber \\
&-&\frac{ig_{BF}}{\hbar}\sum_m \int_0^tdt'\int
d^3r\,e^{-i\varepsilon_n(t-t')/\hbar}\Xi ({\bf r},t')
\left[\alpha_{m \uparrow}(t')(u_n^*u_m-v_n^*v_m)
-\alpha^{\dagger}_{m \downarrow}(t')(u_n^*v_m^*+v_n^*u_m^*) \right] ,  \\
\alpha_{n \downarrow}^{\dagger}(t)&=&\alpha_{n
\downarrow}^{\dagger}(0)e^{i\varepsilon_nt/\hbar} \nonumber \\
&+&\frac{ig_{BF}}{\hbar}\sum_m \int_0^tdt'\int
d^3r\,e^{i\varepsilon_n(t-t')/\hbar}\Xi ({\bf r},t')
\left[\alpha_{m \uparrow}(t')(u_nv_m+v_nu_m) +\alpha^{\dagger}_{m
\downarrow}(t')(u_nu_m^*+v_nv_m^*) \right].
\end{eqnarray}
\label{quas1}
\end{subequations}
\end{widetext}
The eigenenergies for the quasiparticles, $\varepsilon_n$, are
obtained from the Bogoliubov-de Gennes equations
\begin{eqnarray}
\varepsilon_nu_n({\bf r}) &=&\hat{h}_F'u_n({\bf r})+\Delta({\bf
r})v_n({\bf r}), \nonumber \\
\varepsilon_nv_n({\bf r}) &=&-\hat{h}_F'v_n({\bf r})+\Delta^*({\bf
r})u_n({\bf r}) \nonumber.
\end{eqnarray}
Following the same strategy as previously, the quasiparticle
operators inside the integrals of Eqs.~(\ref{quas1}) are first
replaced by their free evolution values for $g_{BF}=0$, and the
density fluctuations of the Fermi gas are calculated using
Eqs.~(\ref{can1}) and Eq.~(\ref{rho1}). However, instead of
$|F\rangle$, here the expectation value is calculated with respect
to the BCS ground state, $|\Phi_0\rangle$. We recall that
$|\Phi_0\rangle$ is the vacuum state for the quasiparticles so
that only terms of the form $\langle
\Phi_0|\alpha_{n\sigma}\alpha^{\dagger}_{n'\sigma'}|\Phi_0\rangle=
\delta_{n,n'}\delta_{\sigma,\sigma'}$ give a non-vanishing
contribution to $\delta\rho_{\sigma}$. Carrying out this
procedure, one arrives at equations for the boson density
fluctuations that have the exact same form as Eq. (\ref{bb})
except that ${\cal J} ({\bf r},{\bf r'},t-t') $ is replaced by the
expression
\begin{equation}
\tilde{{\cal J}} ({\bf r},{\bf r'},t-t')
=\frac{1}{2}\sum_{n,m}\left[F_{n,m}({\bf r},t'-t)F^*_{m,n}({\bf
r}',t-t')-c.c \right], \nonumber
\end{equation}
where $F_{n,m}({\bf r},t)$ is defined as
\begin{equation}
F_{n,m}({\bf r},t)=\left[u_n^*({\bf r})v_m^*({\bf r})+v_n^*({\bf
r})u_m^*({\bf r})\right]e^{-i\varepsilon_nt/\hbar}.
\end{equation}
Using $\tilde{{\cal J}}({\bf r},{\bf r'},t-t')$ in Eq. (\ref{bb})
gives the effect on the bosons of density fluctuations in the
Fermi gas resulting from the creation of pairs of BCS
quasiparticles.

Again, we consider the specific case of a uniform system of volume
$V$. In this case the quasiparticle amplitudes are plane waves
\[ u_n({\bf r})=\frac{1}{\sqrt{V}}U_{{\bf k}}e^{i{\bf k\cdot r}}
,\; v_n({\bf r})=\frac{1}{\sqrt{V}}V_{{\bf k}}e^{i{\bf k\cdot
r}},\] and the energies of the quasiparticles are given by
\[ \varepsilon_{{\bf k}}=\sqrt{E_{{\bf k}}^2+\Delta^2} .\]
The order parameter, $\Delta ({\bf r})=\Delta=(g_F/V)\sum_{{\bf
k}}U_{{\bf k}}V_{{\bf k}}$, is a constant. As mentioned before,
the Hartree-Fock energy is absorbed into the definition of the
Fermi energy, $E_F=\mu_F-g_{BF}n_B+g_Fn_F$. The amplitudes are
most easily expressed in terms of the angle $\theta_{{\bf k}}$
defined by
\[U_{{\bf k}}=\cos(\theta_{{\bf k}}/2),\;V_{{\bf k}}=\sin(\theta_{{\bf
k}}/2),\] where $\tan\theta_{{\bf k}}=\Delta/E_{{\bf k}}$, which
along with $\varepsilon_{{\bf k}}$, is obtained from the solution
of the Bogoliubov-de Gennes equations.

By following the procedure of Sec. II, we obtain solutions for the
Laplace transforms of the $-{\bf k}$ component of the boson
density fluctuation that are identical to Eqs.~(\ref{as}), except
that the density-density response of the ideal Fermi gas,
$\ell_k(s)$, is replaced by the density-density response of the
BCS state, $\tilde{\ell}_k(s)$. It is given by
\begin{widetext}
\begin{eqnarray}
\tilde{\ell}_k(s) &=&\frac{i(2\pi)^3}{V} L\left\{\sum_{{\bf k_m}}
\left( e^{i(\varepsilon_{{\bf k+k_m}}+\varepsilon_{{\bf
k_m}})t/\hbar} -c.c \right)\sin^2\left[\frac{1}{2}(\theta_{{\bf
k+k_m}}+\theta_{{\bf k_m}})\right] \right\},\nonumber \\
&=& \int d{\bf k_m} \,\left[ \frac{1} {-is-(\varepsilon_{{\bf
k+k_m}}+\varepsilon_{{\bf k_m}})/\hbar}
-\frac{1}{-is+(\varepsilon_{{\bf k+k_m}}+\varepsilon_{{\bf
k_m}})/\hbar} \right] \sin^2\left[\frac{1}{2}(\theta_{{\bf
k+k_m}}+\theta_{{\bf k_m}})\right]. \label{dens-dens-sup}
\end{eqnarray}
\end{widetext}
It is easy to show that for $\Delta=0$, one recovers the results
of Sec.~II.

Physically, the poles of $\tilde{\ell}_k(s)$ correspond to the
energies required to create a pair of quasiparticles with momentum
${\bf k+k_m}$ and ${\bf k_m}$, just as was the case for
$\ell_k(s)$. When comparing Eqs. (\ref{dens-dens}) and
(\ref{dens-dens-sup}), we observe that the Heaviside step
functions are replaced by $\sin^2\left[(\theta_{{\bf
k+k_m}}+\theta_{{\bf k_m}})/2\right]$. Physically, this accounts
for the lack of a sharp Fermi surface in the BCS state. We recall
that $\tilde{\ell}_k(s)$ accounts for the boson-fermion
interaction which results from a coupling between the local
densities of the two gases. In ${\bf k}$-space, the interaction
takes the form of a coupling between the $-{\bf k}$ component of
the boson density fluctuation and the ${\bf k}^{th}$ component of
the fermion density, $\rho_{\bf k}$. In terms of the BCS
quasiparticle operators, $\rho_{\bf k}$ has the form
\cite{tinkham},
\begin{widetext}
\begin{equation}
 \rho_{\bf k}= \sum_{{\bf k_m}}\left[ \left(U_{{\bf
k+k_m}}V_{{\bf k}}+U_{{\bf k}}V_{{\bf k+k_m}} \right) \left(
\alpha_{{\bf k+k_m} \uparrow}^{\dagger} \alpha_{{\bf
k}\downarrow}^{\dagger} +\alpha_{{\bf k+k_m}
\downarrow}\alpha_{{\bf k}\uparrow} \right) +\left(U_{{\bf
k+k_m}}U_{{\bf k}}-V_{{\bf k}}V_{{\bf k+k_m}}\right) \left(
\alpha_{{\bf k+k_m} \uparrow}^{\dagger} \alpha_{{\bf k}\uparrow}
+\alpha_{{\bf k} \downarrow}^{\dagger}\alpha_{{\bf
k+k_m}\downarrow} \right) \right].
\end{equation}
\end{widetext}
When $\rho_{\bf k}$ acts on $|\Phi_0\rangle$, only the first term,
which creates two quasiparticles, gives a nonzero contribution.
Consequently, $|U_{{\bf k+k_m}}V_{{\bf k}}+U_{{\bf k}}V_{{\bf
k+k_m}}|^2=\sin^2\left[(\theta_{{\bf k+k_m}}+\theta_{{\bf
k_m}})/2\right]$ is the probability to create a pair of
quasiparticles as a result of a fluctuation in the $-{\bf k}$
component of the density of the Bose gas.

Using $s=i\omega+0^+$, the imaginary part of $\tilde{\ell}_k(s)$
is
\begin{eqnarray}
{\rm Im} [\tilde{\ell}_{k}(\omega)] &= & \pi \int d{\bf k_m} \,
\sin^2\left[(\theta_{{\bf
k+k_m}}+\theta_{{\bf k_m}})/2\right] \nonumber \\
& & \times \,\delta\left( \omega-\varepsilon_{{\bf k_m}+{\bf
k}}/\hbar- \varepsilon_{{\bf k_m}}/\hbar \right).
 \label{im2}
\end{eqnarray}
In order for the BEC to be stable, ${\rm Im}
[\tilde{\ell}_{k}(\omega)]=0$, the energy of a density fluctuation
in the BEC must satisfy
\begin{equation}
\hbar\omega<2\Delta, \label{cri2}
\end{equation}
where $2\Delta$ is the minimum energy needed to create a pair of
quasiparticles in the Fermi gas. This condition is qualitatively
different from (\ref{cri}), which imposed a lower limit on the
energy of the condensate excitations. The presence of Cooper
pairing acts to stabilize the low energy excitations of the
condensate. However, for shorter wavelength excitations, the
quasiparticle energy can exceed 2$\Delta$. This condition limits
the wave number of the stable excitation to be below some maximum
number which can be estimated using the unperturbed Bogoliubov
spectrum as $k_{max}\approx 2\Delta/(\hbar c_0)$ .

\begin{figure}
\begin{center}
    \includegraphics*[width=0.95\columnwidth,
height=0.7\columnwidth]{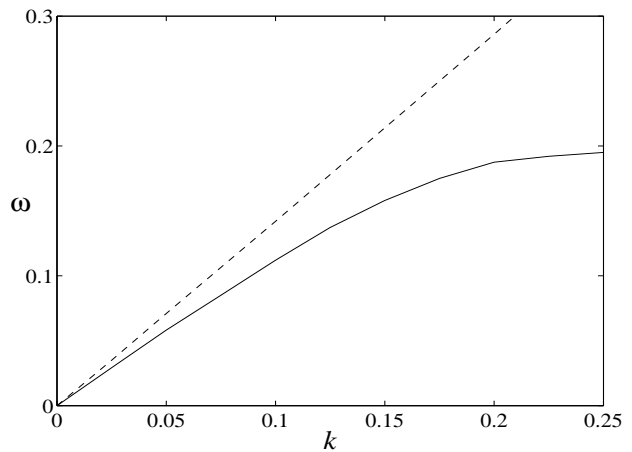} \vspace{3 mm} \caption{Phonon
spectrum of a boson-fermion mixture. The Fermi gas is in a
superfluid state. $a_{BF}=0$ for the dashed line, and $a_{BF}=0.1$
for the solid line. Other parameters are $a_B=0.04$, $n_B=1$ and
$\Delta =0.1$. Same units as in Fig.~\ref{fig1}.} \label{fig3}
\end{center}
\end{figure}
Due to the form of ${\rm Re} [\tilde{\ell}_{k}(\omega)]$, the
dispersion relation, $\omega(k)$, in the stable regime must be
evaluated numerically. Fig.~\ref{fig3} shows $\omega(k)$ for the
BCS state of the Fermi gas. In contrast to the normal state for
the fermions, the BCS state results in an $\omega(k)$ that is
reduced below that of the Bogoliubov spectrum of a pure
condensate. Again, this is a result of the stability criterion
(\ref{cri2}), imposed on the energy of the excitation of the
condensate. For the BCS state in the stable regime determined by
(\ref{cri2}), the density-density response function
$\tilde{\ell}_{k}(\omega)$ is always negative. From
Eq.~(\ref{dispersion}), it follows that $\omega(k)$ is reduced in
this case.

It is worth pointing out that when $\Delta \ll E_F$ (which is
usually the case), one can show that $\tilde{\ell}_{k}(\omega)$ in
the $k \rightarrow 0$ and $\omega=0$ static limit yields
$\tilde{\ell}_{k}(0)\approx -8\pi m_F k_F/ \hbar$, which agrees
with the analytic result for the normal state of the Fermi gas
[see Eq.~(\ref{limit})]. This indicates that the mechanical
stability of the Bose-Fermi mixture does not depend on the state
of the Fermi gas.

\section{Conclusions}
In this paper, by extending the standard Bogoliubov linearization
procedure, we have analyzed the excitations of a weakly
interacting Bose-Einstein condensate coupled to a degenerate Fermi
gas at zero temperature. We derived general expressions for the
excitations of the condensate in the presence of a Fermi gas that
are valid for abitrary spatial geometries. When we specialized our
results to the case of a spatially homogenous system, it was found
that the quasiparticle spectrum for the condensate exhibits a
dynamical instability due to the coupling between the Bose and
Fermi gases. The instability corresponds to the resonant exchange
of energy and momentum between the bosonic quasiparticle and pairs
of quasiparticle excitations in the Fermi gas. In the stable
regime, quantum fluctuations in the density of the Fermi gas
modify the quasiparticle spectrum of the BEC. In the long
wavelength limit, the speed of sound in the BEC is increased
(decreased) when the Fermi gas is in the normal (superfluid) state
as compared to the Bogoliubov speed of sound for a pure weakly
interacting condensate. This difference arises from the different
stability criteria [see Eqs.~(\ref{cri}) and (\ref{cri2})] which
are determined by the nature of the resonant coupling between the
bosons and fermions in the mixture.

This paper lays the groundwork for the study of nonlinear
wave-mixing between degenerate beams of bosons and fermions.
Future work will extend the results obtained here for the
equilibrium state of coupled Bose-Fermi gases to the
nonequilibrium mixing of bosonic and fermionic matter waves. In
contrast to the equilibrium case, where the instability signals
the existence of a new ground state of the system, the existence
of an instability in the nonequilibrium wave-mixing indicates
exponential growth in one of the matter wave modes.

In the current work, we have focused on the quasiparticle
excitation spectrum of the bosons in the mixture. Future work
should include the study of the induced fermion-fermion coupling
due to their interaction with bosons. This will shed light on the
long-sought goal of inducing Cooper pairing of fermions using
bosonic atoms \cite{viverit,bijlsma}.

\acknowledgments This work is supported in part by the US Office
of Naval Research under Contract No. 14-91-J1205, by the National
Science Foundation under Grants No. PHY98-01099 and PHY0098129, by
the US Army Research Office, by NASA Grant No. NAG8-1775, and by
the Joint Services Optics Program.

\section{Appendix: Ground State with Correlated Bose-Fermi Densities}
In this appendix we examine the ground state energy of a BEC
collisionally coupled to a normal Fermi gas in a box of volume
$V$. We show that a variational ground state wave function with a
finite probability amplitude for excitations with opposite
momentum in the BEC and the Fermi gas can result in an energy that
is lower than that of the ground state used in Section II. The
variational wave function we propose in this appendix is not
necessarily the true ground state of the system, but instead
provides an indication of the possible form of the ground state
that the system evolves to in the presence of a dynamical
instability.

The Hamiltonian, $\hat{H}'$ for the Bose-Fermi mixture written in
a plane wave basis has the form (note that in this appendix we do
not use the grand canonical Hamiltonian),
\begin{equation}
\hat{H}'=\hat{H}_T+\hat{H}_{BB}+\hat{H}_{BF},
\end{equation}
where
\begin{eqnarray*}
\hat{H}_T &=& \sum_{{\bf k}} \left(
\frac{\hbar^2k^2}{2m_B}\eta^{\dagger}_{{\bf k}}\eta_{{\bf k}}
+\sum_{\sigma}\frac{\hbar^2k^2}{2m_F}a^{\dagger}_{{\bf
k}\sigma}a_{{\bf k}\sigma} \right),\\
\hat{H}_{BB} &=&\frac{g_B}{2V}\sum_{{\bf k,k',q}}
\eta^{\dagger}_{{\bf k+q}}\eta^{\dagger}_{{\bf k'-q}}\eta_{{\bf
k'}}\eta_{{\bf k}},\\
\hat{H}_{BF} &=&\frac{g_{BF}}{V}\sum_{\bf q}\zeta_{\bf
q}\rho^{\dagger}_{{\bf q}}.
\end{eqnarray*}
$\hat{H}_T$ represents the kinetic energy for the Bose and Fermi
gases while $\hat{H}_{BB}$ is the collisional interaction between
bosons. The boson-fermion interaction, $\hat{H}_{BF}$, has been
expressed in terms of the operators for the ${\bf q}^{th}$ Fourier
component of the fermion density,
\begin{equation}
\rho_{\bf q}=\sum_{{\bf k'}\sigma}a^{\dagger}_{{\bf
k'+q}\sigma}a_{{\bf k'}\sigma}
\end{equation}
and the boson density,
\begin{equation}
\zeta_{\bf q}=\sum_{{\bf k'}}\eta^{\dagger}_{{\bf k'+q}}\eta_{{\bf
k'}}.
\end{equation}
Note that $\rho_{-{\bf q}}=\rho^{\dagger}_{\bf q}$ and similarly,
$\zeta_{-{\bf q}}=\zeta^{\dagger}_{{\bf q}}$. In contrast to
Sections II and III, where the ${\bf k}=0$ condensate mode for the
bosons was treated as a $c$-number, we now retain the operator
dependence for the ${\bf k}=0$ mode, $\eta_{0}$, in the
Hamiltonian. As in Sec. II, we neglect the direct fermion-fermion
interaction.

In Sec. II, the equations of motion for the density fluctuations
were linearized around the $T=0$ ground state for the
non-interacting Bose and Fermi gases, $ |\Psi_N\rangle$, which was
implicitly assumed to remain a stable ground state for the
interacting Bose-Fermi system. However, the existence of the
dynamical instability indicates that $|\Psi_N\rangle$ is actually
not stable and that there exists a ground state with a lower
energy than $|\Psi_N\rangle$. The expectation value of the
Hamiltonian with respect to $|\Psi_N\rangle$, $E_N=\langle
\Psi_N|\hat{H}'|\Psi_N\rangle$, is easily found to be,
\begin{equation}
E_N=2\sum_{k\leq k_F}\frac{\hbar^2k^2}{2m_F}+\frac{g_B}{V}\left(
N_B^2-N_B \right)+\frac{2g_{BF}}{V}N_BN_F.
\end{equation}
$|\Psi_N\rangle$ is equivalent to the ground state used in
previous investigations of Bose-Fermi mixtures, in the sense that
$|\Psi_N\rangle$ does not include any quantum correlations between
the bosons and fermions, i.e. $|\Psi_N\rangle$ factorizes in to
the product of the wave functions for the condensate and the ideal
Fermi gas.

Any excitation with finite momentum in the two gases will increase
the total kinetic energy, $\hat{H}_T$, but may result in a
decrease in the interaction energy between the bosons and
fermions. For example, consider the wave function,
\begin{eqnarray*}
|\Psi_D\rangle &=&\left( u_{\bf q}+v_{\bf q}\kappa^{-1}_{\bf q}
\eta^{\dagger}_{\bf q}\eta_0
\rho^{\dagger}_{{\bf q}}/\sqrt{N_B} \right)|\Psi_N \rangle \\
&=&\left( u_{\bf q}+v_{\bf q}\kappa^{-1}_{\bf q} \zeta_{\bf q}
\rho_{-{\bf q}}/\sqrt{N_B} \right)|\Psi_N \rangle,
\end{eqnarray*}
where $|u_{\bf q}|^2+|v_{\bf q}|^2=1$ and
\begin{equation}
\kappa_{\bf q}^2=\langle \Psi_N|\rho_{\bf q}\rho^{\dagger}_{\bf
q}|\Psi_N \rangle=2 \sum_{\bf k}\Theta(k_F-k)\Theta(|{\bf
k-q}|-k_F).\nonumber
\end{equation}
The action of $\zeta_{\bf q} \rho_{-{\bf q}}$ on $|\Psi_N\rangle$
is to create a state with a density fluctuation of momentum $\hbar
{\bf q}$ in the BEC along with a density fluctuation of momentum
$-\hbar {\bf q}$ in the Fermi gas. It is easy to show that, just
like $|\Psi_N\rangle$, $|\Psi_D\rangle$ corresponds to a spatially
uniform state with densities $N_B/V$ and $2N_F/V$ for the Bose and
Fermi gases, respectively. The difference between $|\Psi_D\rangle$
and $|\Psi_N\rangle$ is made manifest in the correlation between
the boson and fermion densities,
\begin{eqnarray}
\langle \Psi_N|\hat{\psi}_B^{\dagger}\hat{\psi}_B({\bf r})
\sum_{\sigma}\hat{\psi}_{\sigma}^{\dagger}\hat{\psi}_{\sigma}({\bf
r'}) |\Psi_N\rangle
= \frac{N_B}{V}\frac{2N_F}{V}, & \label{ddN} \\
\langle \Psi_D|\hat{\psi}_B^{\dagger}\hat{\psi}_B({\bf r})
\sum_{\sigma}\hat{\psi}_{\sigma}^{\dagger}\hat{\psi}_{\sigma}({\bf
r'}) |\Psi_D\rangle
=\frac{N_B}{V}\frac{2N_F}{V} & \nonumber \label{ddD} \\
+\frac{2\sqrt{N_B}\kappa_{\bf q}}{V^2} |u_{\bf q}||v_{\bf
q}|\cos[{\bf q}\cdot ({\bf r}-{\bf r'})+\gamma],
\end{eqnarray}
where $\gamma=\arg(u^*_{\bf q}v_{\bf q})$ is the relative phase
between $u_{\bf q}$ and $v_{\bf q}$. The density-density
correlation for $|\Psi_D\rangle$ depends on the quantum
coherences, $u_{\bf q}^*v_{\bf q}$, and can be made larger or
smaller than (\ref{ddN}) by varying $\gamma$. This can be used to
lower the boson-fermion interaction energy since it is
proportional to the spatially integrated density-density
correlation between the two gases for ${\bf r}={\bf r'}$.

The difference in the energy of the two states is
\begin{eqnarray}
\Delta E&=&\langle \Psi_D|\hat{H}'|\Psi_D\rangle - E_N  \nonumber \\
&=&E_1(q)v_{\bf q}^2+\frac{1}{2}E_2(q) \left( v_{\bf q}u^*_{\bf
q}+v^*_{\bf q}u_{\bf q} \right) .\nonumber
\end{eqnarray}
$E_1(q)$ represents the increase in the kinetic energy and the
mean field energy of the condensate,
\begin{equation}
E_1(q)=\frac{\hbar^2q^2}{2m_B}+2 \kappa_{\bf q}^{-2} \sum_{k\leq
k_F}^{|{\bf k-q}|>k_F} \left(
\frac{\hbar^2q^2}{2m_F}-\frac{\hbar^2{\bf k \cdot q}}{m_F} \right)
+2g_B\frac{N_B}{V},\nonumber
\end{equation}
while $E_2(q)$ is due to the boson-fermion interaction,
\begin{equation}
E_2(q)=2g_{BF}\sqrt{N_B}\kappa_{\bf q}/V.\nonumber
\end{equation}
Now let $v_{\bf q}=\sin \phi_{\bf q}$ and $u_{\bf q}=\pm\cos
\phi_{\bf q}$ for $0\leq \phi_{\bf q} \leq \pi/2$. For $u_{\bf
q}$, we use the upper sign for $g_{BF}<0$ and the lower sign for
$g_{BF}>0$. By minimizing $\Delta E$ with respect to $\phi_{\bf
q}$ one finds that the minimum value of the energy difference is
\begin{equation}
\Delta E_{min}=\frac{1}{2}\left[E_1(q)-\sqrt{E_1(q)^2+E_2(q)^2}
\right],\nonumber
\end{equation}
which occurs when $\tan 2\phi_{\bf q}=|E_2(q)|/E_1(q)$. Note that
$\Delta E_{min}<0$ for all values of ${\bf q}$, which indicates
that quantum correlations between the densities of the two gases
can lower the energy. Consequently, $|\Psi_N\rangle$ is not the
true ground state of the Bose-Fermi mixture. To find the true
ground state, $|\Psi_D\rangle$ would have to be extended to treat
density fluctuations in all modes. This is a non-trivial task that
will be the subject of future work.

We want to stress here that even though $|\Psi_N\rangle$ is not
the lowest energy state, this does not necessarily mean that it is
{\em dynamically unstable}. The relationship between the states
$|\Psi_D\rangle$, $|\Psi_N\rangle$ and the dynamical stability
condition, ${\rm Im} [\ell_{k}(\omega)]=0$, can be understood in
the following manner. Suppose we start with the BEC and Fermi gas
in separate unconnected boxes so that the quantum states for the
Bose-Fermi system is factorizable into the product of the BEC
state and the state of an ideal Fermi gas at $T=0$, namely
$|\Psi_N\rangle$. If we then bring the two gases together in the
same box so that they can interact, then $|\Psi_N\rangle$ will
evolve into an entangled state with a form similar to
$|\Psi_D\rangle$ provided that ${\rm Im} [\ell_{q}(\omega(q))]\neq
0$, where $\hbar\omega(q)$ is the energy of the density
fluctuation in the BEC. This is because (a) correlations between
the density fluctuations only develop if there is a coupling
between the fluctuations in the two gases; and (b) since ${\rm Im}
[\ell_{q}(\omega(q))]$ is proportional to the dynamic structure
factor for the Fermi gas, it measures the strength of the resonant
coupling between the two gases. If ${\rm Im}
[\ell_{q}(\omega(q))]=0$, the density fluctuations in the two
gases with momentum $\pm\hbar {\bf q}$ are uncoupled and there is
no way to generate any quantum correlations between the two gases,
thereby lowering the energy of the system.

This argument can be made quantitative if we evaluate the state of
the Bose-Fermi system to first order in pertubation theory. If we
work in the interaction representation where
\begin{equation}
\hat{H}_{BF}(t)=e^{i(\hat{H}_T+\hat{H}_{BB})t/\hbar}\hat{H}_{BF}
e^{-i(\hat{H}_T+\hat{H}_{BB})t/\hbar},\nonumber
\end{equation}
then the state of the system at time $t$ starting from the ground
state of the uncoupled system is, to first order in $g_{BF}$,
\begin{equation}
|\Psi(t)\rangle=\left(1-i\hbar^{-1}\int_{-\infty}^{t}dt'\hat{H}_{BF}(t')
\right)|\Psi_N\rangle. \nonumber
\end{equation}
By letting $t\rightarrow \infty$ we obtain,
\begin{widetext}
\begin{equation}
|\Psi (\infty)\rangle=\left(1-2i\pi g_{BF}\hbar^{-1}\sum_{\bf
q}\eta^{\dagger}_{\bf q}\eta_{0}\left[\frac{1}{V}\sum_{{\bf
 k}\sigma}a^{\dagger}_{{\bf k-q}\sigma}a_{{\bf
 k}\sigma}\delta\left(\omega(q)-\hbar({\bf k-q})^2/2m_F+\hbar{\bf
 k}^2/2m_F\right)\right]\right)|\Psi_N\rangle, \label{perturb2}
\end{equation}
\end{widetext}
which has a form similar to that of $|\Psi_D\rangle$ generalized
to include all modes for the coupled density fluctuations. The
delta function implies that correlations are dynamically generated
between $\zeta_{{\bf q}}$ and those components of $\rho_{\bf q}$
that conserve energy in the $t\rightarrow \infty$ limit. Note that
$\hbar\omega(q)$ in Eq. (\ref{perturb2}) is equal to the energy of
a Bogoliubov quasipaticle in the uncoupled BEC since dispersive
effects due to the Bose-Fermi coupling are higher order in
$g_{BF}$. From $|\Psi(\infty)\rangle$, one sees that the
interaction between the bosons and fermions will naturally lead to
an entangled state provided the term in brackets is nonzero. It is
easy to see from Eq.~(\ref{im}) that
\begin{eqnarray}
\left[\sum_{{\bf
 k}\sigma}a^{\dagger}_{{\bf k-q}\sigma}a_{{\bf
 k}\sigma}\delta\left(\omega(q)-\frac{\hbar({\bf k-q})^2}{2m_F}+\frac{\hbar{\bf
 k}^2}{2m_F}\right)\right]
 |\Psi_N\rangle=0 ,\nonumber
\end{eqnarray}
unless ${\rm Im} [\ell_{-q}(\omega(-q))]\neq0$. Note that for an
isotropic system, ${\rm Im} [\ell_{-q}(\omega(-q))]={\rm Im}
[\ell_{q}(\omega(q))]$.

To summarize, we have shown that the existence of a dynamical
instability indicates that entanglement between the Bose and Fermi
systems can be dynamically generated by $\hat{H}_{BF}$ starting
from the factorizable state $|\Psi_N\rangle$. In this case, a
variational wave function such as $|\Psi_D\rangle$ that involves
an entanglement between density fluctuations in the two gases with
opposite momenta can have lower energy than $|\Psi_N\rangle$.

\end{document}